\documentclass[12pt]{article}
\pdfoutput=1

\newlength{\abstractwidth}
\usepackage{epsf}
\usepackage{color}
\usepackage{graphicx}
\usepackage{tikz}

\usepackage{amsmath}
\usepackage{amssymb}
\usepackage{latexsym}

\renewcommand{\thefootnote}{\fnsymbol{footnote}}
\renewcommand{\thanks}[1]{\footnote{#1}}
\newcommand{\starttext}{
\setcounter{footnote}{0}
\renewcommand{\thefootnote}{\arabic{footnote}}}
\newcommand{\bea}{\begin{eqnarray}}
\newcommand{\eea}{\end{eqnarray}}
\newcommand{\be}{\begin{eqnarray}}
\newcommand{\ee}{\end{eqnarray}}

\def\cD{{\cal D}}

\def\cH{{\cal H}}

\def\cL{{\cal L}}

\def\cO{{\cal O}}

\def\ZZ{{\mathbb Z}}
\def\RR{{\mathbb R}}

\def\CC{{\mathbb C}}

\def\Im{{\rm Im \,}}

\def\half{{1\over 2}}
\def\thalf{{\tfrac{1}{2}}}

\def\a{\alpha}

\def\tet{\vartheta}

\def\no{\nonumber}
\def\sm{\smallskip}

\begin{document}
\starttext
\setcounter{footnote}{0}

\begin{flushright}
2019 April 13 \\
revised 2019 October 4 \\
DAMTP-2019-16 
\end{flushright}

\vskip 0.3in

\begin{center}

{\Large \bf Absence of irreducible multiple zeta-values in\\  melon modular graph functions }

\vskip 0.2in

{\large Eric D'Hoker$^{(a)}$, Michael B. Green$^{(b)}$} 

\vskip 0.15in

{ \sl  (a) Mani L. Bhaumik Institute for Theoretical Physics}\\
{\sl  Department of Physics and Astronomy}\\
{\sl University of California, Los Angeles, CA 90095, USA}

\vskip 0.1in

{ \sl (b) Department of Applied Mathematics and Theoretical Physics }\\
{\sl Wilberforce Road, Cambridge CB3 0WA, UK}, and \\
{\sl Centre for Research in String Theory, School of Physics, }\\
{\sl Queen Mary University of London, Mile End Road, London, E1 4NS, England}

\vskip 0.15in

{\tt \small dhoker@physics.ucla.edu, M.B.Green@damtp.cam.ac.uk}

\vskip 0.5in

\begin{abstract}
\vskip 0.1in
 
 The expansion of a modular graph function on a torus of modulus $\tau$ near the cusp is given by a Laurent polynomial in $y= \pi \Im (\tau)$ with coefficients that  are rational multiples of  single-valued multiple zeta-values, apart from the leading term whose coefficient is rational and exponentially suppressed terms.  We prove that the coefficients of the non-leading terms in the Laurent polynomial of  the modular graph function $D_N(\tau)$ associated with a melon  graph is  free of irreducible multiple zeta-values and can be written  as a polynomial in odd zeta-values with rational coefficients  for arbitrary $N \geq 0$.    The proof proceeds by expressing a generating function for $D_N(\tau)$  in terms of an integral over the Virasoro-Shapiro closed-string tree amplitude.  

\end{abstract}

\end{center}

\baselineskip=15pt
\setcounter{footnote}{0}

\newpage

A genus-one modular graph function is an $SL(2,\ZZ)$-invariant function on the Poincar\'e upper half plane $\cH$ which is associated with a Feynman graph for a massless scalar field  on a torus  \cite{DHoker:2015wxz}. Modular graph functions arise as the basic building blocks for the coefficients of the effective interactions in a low energy expansion of string theory. One-loop modular graph functions are given in terms of the classic non-holomorphic Eisenstein series, while two-loop modular graph functions have been studied only recently  in \cite{DHoker:2015gmr,DHoker:2017zhq}. In particular,  their Fourier series representation, as well as their Poincar\'e series representation as a sum over cosets $\Gamma _\infty \backslash SL(2,\ZZ)$, are by now explicitly known \cite{DHoker:2019txf}.  The expansion of a generic modular  graph function on a torus with modulus $\tau \in \cH$ near the cusp reduces to a Laurent polynomial in  $1/y$, where $y= \pi \,\Im\tau$, plus exponentially suppressed terms.  The leading term in the Laurent polynomial for a modular graph function of weight $N$ is a rational number multiplying $y^N$ and the coefficients of all  succeeding terms are single-valued multiple  zeta-values.

\sm

The general structure of  modular graph functions with three loops or more is not understood as explicitly, though many systematic results were obtained   in \cite{DHoker:2015gmr,Basu:2015ayg,Zerbini,DHoker:2016mwo,Basu:2016kli,DHoker:2016quv,Kleinschmidt:2017ege,Brown:2017qwo,Brown2}. One exception is the melon modular graph functions $D_N$ of weight $N$ whose Feynman graph is represented in Figure 1, and whose full Laurent polynomial was computed in \cite{Green:2008uj} in terms of multiple zeta-values. The goal of this note is to provide a simple proof that the coefficients of the Laurent series of  $D_N$ are actually free of irreducible multiple zeta-values and  given by a polynomial in odd zeta-values only, plus a leading $y^N$ term, both with rational coefficients.\footnote{ MBG is very grateful  to Don Zagier for discussions in 2012 concerning his arguments for the absence of irreducible multiple zeta values in the Laurent polynomial of $D_N$ functions,   although this has not appeared in published form.  We believe that the present proof is significantly simpler and leads to expressions for the Laurent polynomial coefficients that are easier to evaluate.} The full Laurent polynomial for each $D_N$ is given in terms of odd zeta-values  by a fairly simple generating function.

\begin{figure}[h]
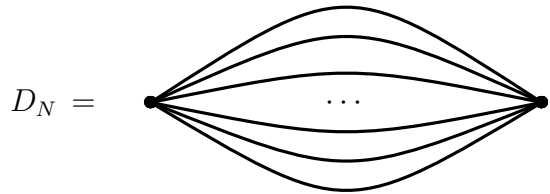

\begin{center}
\tikzpicture[scale=1.3]
\scope[xshift=-5cm,yshift=-0.4cm]
\draw (-1,0) node{$D_N \, =$};
\draw[very thick]   (0,0) node{$\bullet$}  ..controls (2,1.3) .. (4,0) node{$\bullet$} ;
\draw[very thick]   (0,0) node{$\bullet$} ..controls (2,0.9) .. (4,0) node{$\bullet$} ;
\draw[very thick]   (0,0) node{$\bullet$} ..controls (2,0.4) .. (4,0) node{$\bullet$} ;
\draw[very thick]   (2,0) node{$\cdots$} ;
\draw[very thick]   (0,0) node{$\bullet$} ..controls (2,-0.4) .. (4,0) node{$\bullet$} ;
\draw[very thick]   (0,0) node{$\bullet$} ..controls (2,-0.8) .. (4,0) node{$\bullet$} ;
\draw[very thick]   (0,0) node{$\bullet$} ..controls (2,-1.2) .. (4,0) node{$\bullet$} ;
\endscope
\endtikzpicture
\caption{The melon modular graph function $D_N$ has $N$ Green functions joining the points.}
\end{center}
\end{figure}

We shall denote the modulus of the torus $\Sigma$ by $\tau= \tau_1+i \tau_2$ with $\tau_1, \tau_2 \in \RR$ and $\tau_2>0$ and choose a local complex coordinate $z= \a+\beta \tau$ with  $\alpha , \beta \in [-\half, \half]$ and volume form $d^2z = { i \over 2} dz \wedge d\bar z= \tau_2 d\alpha \wedge d \beta$. The modular graph function $D_N$ may be expressed as follows,  
\bea
\label{one}
D_N (\tau) = \int _\Sigma { d^2 z \over \tau_2} \, G(z|\tau)^N
\eea
The scalar Green function $G(z|\tau)$ on $\Sigma$ satisfies the standard Laplace equation with a unit $\delta$-function source at $z=0$ and is given by the following expression (for a review of Riemann surfaces in string theory and explicit formulas, see for example  \cite{DHoker:1988pdl}),
\bea
\label{two}
G(z|\tau) = - \ln \left | { \tet_1(z|\tau) \over \eta (\tau)} \right |^2 + 2 \pi \tau_2 \beta ^2 
\eea
where $\eta$ is the Dedekind eta-function and $\tet_1$ the Jacobi theta-function. 
Equivalently the Green function may be expressed in a Fourier series in the variable $\alpha$, 
\bea
\label{three}
G(z|\tau) = 2 \pi \tau_2 (\beta^2 - |\beta| +\tfrac{1}{6} ) + \sum _{m \not=0} \sum _k {1 \over |m|} e^{ 2 \pi i m ( \alpha + \beta \tau_1 + k \tau_1)- 2 \pi \tau_2 |m(k+\beta)|}
\eea
The Green function is normalized so that $D_1=\int _\Sigma d^2 z \, G(z|\tau)=0$. The full  Laurent polynomial of $D_N(\tau)$ in terms of the variable $y=\pi \tau_2$ near the cusp $y \to \infty$ was obtained in \cite{Green:2008uj}  by substituting the expression for $G(z|\tau)$ of (\ref{three}) into (\ref{one}) to obtain, 
\bea
\label{four}
D_N(\tau) & = &  { y^N \over 3^N} {}_2F_1(1,-N;\tfrac{3}{2}; -\tfrac{3}{2}) 
+ \sum _{k=0}^{N-2} \sum_{{k_1, k_2, k_3 \geq 0 \atop k_1+k_2+k_3=k}} { 2(-)^{k_2} N! \, (2k_1+k_2)! (2y)^{k_3-k_1-1} \over 6^{k_2} \, (N-k)! \, k_1! \, k_2 ! \, k_3! }
\no \\ && \hskip 2in  \times 
S(N-k,2k_1+k_2+1) + \cO(e^{-4y})
\eea
where ${}_2F_1$ is the hypergeometric function. 
The coefficients $S(M,N)$ are defined for $M,N\geq 1$ by the following multiple series, 
\bea
S(M,N) = \sum _{{m_r \not=0 \atop r=1, \cdots, M}} { \delta ( \sum _r m_r) \over |m_1 \cdots m_M|(|m_1|+\cdots + |m_M|)^N}
\eea
Zagier showed (Appendix A of  \cite{Green:2008uj})  that $S(M,N)$ is expressible as a linear combination of multiple zeta-values, 
\bea
S(M,N) = \sum _{{a_1, \cdots, a_r \in \{ 1,2\} \atop a_1+\cdots +a_r=M-2}} M! ~ 2^{2r+2-M-N} ~ \zeta (N+2, a_1, \cdots, a_r)
\eea
where a multiple zeta-value of depth $\ell$ is defined by,
\bea
\zeta (s_1, \cdots, s_\ell) = \sum _{n_1 > n_2 > \cdots > n_\ell  \geq 1} { 1 \over n_1 ^{s_1} \cdots n_\ell ^{s_\ell}}
\eea
It was conjectured in \cite{Green:2008uj}, on the basis of results obtained for low values of $N$,  that the coefficients of the Laurent expansion of $D_N$ are actually free of irreducible multiple zeta-values (namely those which cannot be expressed as a polynomial in zeta-values).  Since Zerbini's explicit calculations \cite{Zerbini} of the Laurent polynomials of various modular graph functions do exhibit irreducible zeta-values, the conjecture on $D_N$ is non-trivial and implies an arithmetic simplicity of the $D_N$ functions not shared by general modular graph functions. Zagier has argued in an unpublished paper that the conjecture holds,  but his procedure is quite involved \cite{Zagier} and appears to follow a different path from the simple proof of the theorem below that will be presented in this note.  

\vskip 0.1in

{\bf Theorem 1}  {\sl The Laurent polynomial, in $y=\pi \tau_2$ at the cusp $y \to \infty$, of the modular graph function $D_N(\tau)$ satisfies the following properties,}
\begin{enumerate}
\itemsep=0in
\item {\sl it is free of irreducible multiple zeta-values;}
\item {\sl the coefficient of its leading monomial $y^N$ is  rational, while the coefficient of each one of its sub-leading monomials is a polynomial in odd zeta-values  with rational coefficients;} 
\item {\sl  it is homogeneous  in the weight and of total weight $N$, provided we assign weight $n$ to $\zeta(n)$ and weight 1 to $y$.}
\end{enumerate}

To prove the theorem, we use a generating function for the modular graph functions $D_N$, 
\bea
\cD(s|\tau) = \sum _{N=0}^\infty { s^N \over N !} D_N (\tau) = \int _\Sigma {d^2 z \over \tau_2} \, e^{s \, G(z|\tau)}
\label{cddef}
\eea
Having assigned weight $N$ to the modular graph function $D_N(\tau)$ it is natural to assign weight $-1$ to the variable $s$ so that the generating function $\cD(s|\tau)$ has weight zero. We shall use equation (\ref{two}) for the Green function $G(z|\tau)$ and express $\tet_1(z|\tau)$ and $\eta(\tau)$ in terms of their respective infinite product formulas to obtain, 
\bea
\label{thetaprod}
{ \tet_1(z|\tau) \over \eta (\tau)}  = i \, e^{i \pi \tau/6} \left ( e^{ i \pi z} - e^{-i \pi z} \right ) 
\prod _{n=1}^\infty \left (1 - e^{ 2 \pi i n \tau + 2 \pi i z  } \right )  \left (1 - e^{2 \pi i n \tau  -2 \pi i z } \right ) 
\eea
Since the Green function $G(z|\tau)$ and the domain of integration $\Sigma = \{ \alpha, \beta  \in [- \thalf, \thalf] \}$ are invariant under $z \to -z$, we may restrict the integration to  $\alpha \in [-\half, \half]$ and $ \beta \in [ 0, \half]$ upon including an overall factor of 2, so that we have, 
\bea
\cD(s|\tau) = 2  \int _0 ^\half d \beta \, \int _{-\half} ^\half d \alpha \, e^{s \, G(z|\tau)}
\label{cddef1}
\eea
In the  domain $\alpha  \in [- \thalf, \thalf], \beta \in [0,\half]$  the contribution to (\ref{thetaprod})  from the infinite product in $n$ equals 1 up to terms that are exponentially suppressed in $\tau$ and of order $\cO(e^{- \pi \tau_2})$, uniformly throughout $\Sigma$. As a result, the Green function in (\ref{cddef1})  may be simplified as follows,
\bea
\label{Green}
G(z|\tau) = { \pi \tau _2 \over 3} + 2 \pi \tau _2 (\beta^2-\beta)  - \ln |1-e^{2 i \pi (\alpha + \tau_1 \beta) - 2 \pi \tau_2 \beta } |^2  + \cO(e^{- \pi \tau_2})
\eea
uniformly in the domain $\alpha \in [- \thalf, \thalf], \beta \in [0,\half]$, and the generating function reduces to, \footnote{Note that the terms of order $\cO(e^{- \pi \tau_2})$ in the Green function cancel upon integration over $\alpha$, so that the leading exponential terms that are being neglected are of order $\cO(e^{- 2 \pi \tau_2})$.}
\bea
\cD(s|\tau) = 2 \, e^{ \pi s \tau_2 /3}  \int _0 ^\thalf d \beta \, \int _{-\half} ^\half d \alpha \, 
e^{ 2 \pi s \tau_2 (\beta^2 -  \beta)} 
\left |1-  e^{ 2 \pi i (\alpha + \tau_1 \beta) - 2 \pi  \beta \tau_2}  \right |^{-2s} + \cO(e^{-2 \pi \tau_2})
\eea
Changing integration variables $(\alpha , \beta)  \to (\alpha - \tau_1 \beta, \beta) $ and using the periodicity of the integrand and integration domain in $\alpha$ with period 1, we establish that all dependence on $\tau_1$ cancels out of the generating function $\cD(s|\tau)$, up to exponentially suppressed terms which do not contribute to the Laurent polynomial in $\tau_2$ of $D_N(\tau)$, and we obtain,   
\bea
\cD(s|\tau) = 2 \, e^{ \pi s \tau_2 /3}  \int _0 ^\thalf d \beta \, \int _0^1 d \alpha \, 
e^{ 2 \pi s \tau_2 (\beta^2 -  \beta)} 
\left |1-  e^{ 2 \pi i \alpha - 2 \pi  \beta \tau_2}  \right |^{-2s} + \cO(e^{-2 \pi \tau_2})
\eea
Next, we isolate the contribution in which the absolute value is set to 1, 
\bea
\cD(s|\tau)= \cD_0(s|\tau) + \cD_1(s|\tau) 
\eea
where $\cD_0$ is the generating function of the leading term in $y=\pi \tau_2$ familiar from (\ref{four}),
\bea
\cD_0(s|\tau) = 2   \int _0 ^\thalf d \beta \, 
e^{ 2 \pi s \tau_2 (\beta^2 -  \beta +\tfrac{1}{6} )} 
= \sum _{N=0}^\infty { s^N y^N \over 3^N N!}  \, {}_2F_1(1,-N;\tfrac{3}{2}; -\tfrac{3}{2}) 
\eea
The remaining contribution then takes the following form,
\bea
\cD_1(s|\tau)  =  2 \, e^{  sy /3}  \int _0 ^\thalf d \beta \, \int _0^1 d \alpha \, 
e^{ 2  s y (\beta^2 -  \beta)} 
\left ( \left | 1- e^{ 2 \pi i \alpha -  2 \beta y}  \right |^{-2s} -1 \right ) + \cO(e^{-2 \pi \tau_2})
\eea
Taylor expanding the exponential of the $2 sy \beta^2$ term in the integrand  in powers of $s$,  we find the following representation, 
\bea
\label{cddef2}
\cD_1(s|\tau)  =  2 \, e^{  sy /3} \sum _{k=0} ^\infty { (2 ys)^k  \over k!} 
\int _0 ^\thalf \!\! d \beta  \int _0^1 \!\! d \alpha  \, \beta ^{2k} \,  e^{ - 2  s y  \beta} 
\left ( \left | 1- e^{ 2 \pi i \alpha -  2 \beta y}  \right |^{-2s} -1 \right ) + \cO(e^{-2 \pi \tau_2})
\eea
Observing that, for each value of $k$, the following integral is exponentially suppressed in $\tau_2$,
\bea
\int _\half ^\infty  d \beta \, \int _0 ^1 d \alpha \,  \beta ^{2k} \,  e^{ - 2  s y  \beta} 
\left ( \left | 1- e^{ 2 \pi i \alpha -  2 \beta y}  \right |^{-2s} -1 \right ) = \cO(e^{-2 \pi \tau_2})
\eea 
we may extend the integration domain for $\beta$ in (\ref{cddef2}) to the half line $\beta >0$ since the difference is proportional to the above exponentially suppressed integral, and we find, 
\bea
\label{cddef3}
\cD_1(s|\tau)  =  2 \, e^{  sy /3} \sum _{k=0} ^\infty { (2ys)^k  \over k!} 
\int _0^1 \!\! d \alpha \int _0 ^\infty \! \! d \beta \, \beta ^{2k} \,  e^{ - 2  s y  \beta} 
\left ( \left | 1- e^{ 2 \pi i \alpha -  2 \beta y}  \right |^{-2s} -1 \right ) + \cO(e^{- 2 \pi \tau_2}) 
\eea
Changing variables from $\alpha, \beta $ to $w =  e^{ 2 \pi i \alpha - 2  \beta y}$, the domain of integration for $w$ becomes the unit disc, and we have, 
\bea
\cD_1(s|\tau) = {e^{ s y /3} \over 2 \pi }  \,  \sum _{k=0} ^\infty 
{ s^k \, L_k (s)  \over k! \,  (2 y)^{k+1} } + \cO(e^{- 2 \pi \tau_2})
\label{lkdef}
\eea
where the coefficients $L_k(s)$ are independent of $y$ and given by, 
\bea
L_k (s) = 2 \int _{|w|\leq 1} \, { d^2 w \over |w|^2} \, |w|^s
\Big ( |1-w  |^{-2s}  -1 \Big ) \Big (  \ln |w| \Big )^{2k} 
\eea
The contribution from the first term in the parentheses in the integrand is invariant under $w \to w^{-1}$ for all $k, s$.
Thus, we may complete its $w$-integration into the full complex plane,
\bea
L_k (s) =  \int _\CC \, { d^2 w \over |w|^2} \, |w|^s
 |1-w  |^{-2s}   (  \ln |w| )^{2k} - 
 2 \int _{|w|\leq 1} \, { d^2 w \over |w|^2} \, |w|^s (  \ln |w| )^{2k} 
\eea
Next, we introduce the following generating function for the coefficients $L_k(s)$,
\bea
\cL(s,\xi)  = \sum _{k=0}^\infty { \xi ^{2k} \over (2k)!} L_k(s)
\label{cldef}
\eea
The integral representation for $\cL(s,\xi)$ is derived from the one for $L_k(s)$, 
\bea
\cL(s,\xi)= 
 \int _\CC \, { d^2 w \over |w|^2} \, |w|^{s+\xi}  |1-w  |^{-2s}  - 
  \int _{|w|\leq 1} \, { d^2 w \over |w|^2} \, |w|^s (|w|^\xi + |w|^{-\xi} )
\eea
where we have used the fact that all odd powers of $\xi$ in the first integral vanish since their integrands are odd under $w \to w^{-1}$. The evaluation of the second integral is straightforward while the evaluation of the first integral is familiar from Shapiro's treatment of  the Virasoro-Shapiro amplitude \cite{Shapiro:1970gy},  
\bea
\int _\CC d^2 w \, |w|^{-2-2a} |1-w|^{-2s} 
=  { \pi s \over (s+a) (-a) } \, { \Gamma (1-s) \Gamma (1-a) \Gamma (1+s+a) \over \Gamma (1+s) \Gamma (1+a) \Gamma (1-s-a) }
\eea
Setting $a=-\half(s+\xi)$ we find the following expression for $\cL(s,\xi)$,
\bea
\cL(s,\xi)= { 4 \pi s \over s^2 -\xi^2} \left ( { \Gamma (1-s) \Gamma (1+\half s+ \half \xi) \Gamma (1+\half s- \half \xi) 
\over \Gamma (1+s) \Gamma (1-\half s- \half \xi) \Gamma (1-\half s+ \half \xi) }
-1 \right )
\eea
The function $\cL(s,\xi)$ is even in $\xi$, as expected from its original definition.
It is standard to express the ratio of $\Gamma$-functions in terms of an exponential of odd zeta-values, and we find, 
\bea
\cL(s,\xi)= { 4 \pi s \over s^2 -\xi^2} \left ( \exp \left \{ \sum _{m=1}^\infty { 2 \zeta(2m+1) \over 2m+1} 
\left [ s^{2m+1} - { (s+\xi)^{2m+1} + (s-\xi)^{2m+1} \over 2^{2m+1}} \right ]  \right \} -1 \right )
\label{clres}
\eea

\sm

The  coefficients $L_k(s)$ are recovered  by expanding  the function $\cL(s,\xi)$  given by  (\ref{clres})  in  powers of $\xi$ and using the definition (\ref{cldef}). Substituting the coefficients $L_k(s)$ obtained in this manner into (\ref{lkdef}) and expanding in powers of $s$ provides an efficient practical construction of the Laurent polynomial for the modular graph function $D_N$ for arbitrary $N$.

\sm

 It is evident that the resulting   expressions for $L_k(s)$ and thus for the Laurent polynomial of $D_N$ are free of irreducible multiple zeta-values, thereby proving part 1. of Theorem 1.   
 
 \sm
 
 Furthermore, it follows from (\ref{lkdef}) and (\ref{cddef}) that the coefficients of all the terms in the Laurent polynomial  in  (\ref{four}),  apart from the term of order $y^N$, are polynomials  in  odd zeta-values with rational coefficients, while the coefficient of  $y^N$ is given by  the first term in (\ref{four}), which is a rational number. This  proves part 2. of Theorem 1.
 
 \sm
 
 Finally, assigning weight $-1$ to the parameter $s$ and  weight 0 to the generating function $\cD(s|\tau)$, as we had argued already earlier based on the weight assignment of $D_N(\tau)$,  and further assigning weight $-1$ to the auxiliary variable  $\xi$, we deduce that the weight of $\cL(s,\xi)$ is $2$, so that the weight of the coefficient $L_k(s)$ is $2k+2$. Combining this result with the Laurent expansion in (\ref{lkdef}), and using the standard assignment of weight 1 to $\pi$, then establishes that $D_N(\tau)$ is given by a term in $y^N$ times a rational number plus a Laurent polynomial in $y$ whose coefficients  are polynomials in odd zeta-values with total weight $N$. This proves part 3.  of  Theorem 1.

\vskip 0.2in 

\noindent
{\bf \large Acknowledgments}

\medskip

We thank Don Zagier for sharing his  ideas on $D_N$ functions, which significantly influenced our work.
We gratefully acknowledge  the Niels Bohr International Academy in Copenhagen for the warm hospitality extended to us during the course of this work.  The research of ED  is supported in part by the National Science Foundation under research grant PHY-16-19926.  MBG has been partially supported by STFC consolidated grant ST/L000385/1, by a Leverhulme Emeritus Fellowship, and by a  Simons Visiting Professorship at the NBIA.

\end{document}